\documentclass[twocolumn]{aa}
\input psfig.tex
\usepackage{graphicx}
\usepackage{txfonts}
\def\lsim{\lower.5ex\hbox{$\; \buildrel < \over \sim \;$}}
\def\gsim{\lower.5ex\hbox{$\; \buildrel > \over \sim \;$}}
\def \simeq{\lower.3ex\hbox{$\; \buildrel \sim \over - \;$}}
\def\ch{\lower-0.55ex\hbox{--}\kern-0.55em{\lower0.15ex\hbox{$h$}}}
\def\lh{\lower-0.55ex\hbox{--}\kern-0.55em{\lower0.15ex\hbox{$\lambda$}}}

.tfm

\begin{document}
\title{Mass accretion rate of the galactic  black hole A0620-00 in its quiescent state}
\author{Sabyasachi Pal\inst{1} and Sandip K. Chakrabarti\inst{2,1}}

\offprints{Sabyasachi Pal}

\institute { Centre for Space Physics, Chalantika 43, Garia Station Rd., Kolkata, 700084, India
\email{space\_phys@vsnl.com}
\and
S. N. Bose National Centre for Basic Sciences, Salt Lake, Kolkata, 700098, India
\email{chakraba@bose.res.in}
}

\date{Received ; accepted , }

\markboth{Accretion rate of A0620-00 in Quiescence}{}

\abstract{
Our recent radio observation using the Giant Meter Radio Telescope  (GMRT) 
of the galactic black hole transient  A0620-00 at 1.280GHz revealed a micro-flare 
of a few milli-Jansky. Assuming an equipartition of magnetic energy and 
the gravitational potential energy of accreting matter, it is possible 
to estimate the upper limit of the accretion rate.      
Assuming a black hole mass of $10M_\odot$ residing at the  center, we
find this to be at the most ${\dot M} = (8.5 \pm 1.4) \times 10^{-11} 
M_\odot$ yr$^{-1}$. This is consistent with earlier estimates of accretion rates
based on optical and X-ray observations.
}

\noindent  PUBLISHED IN Astron. Astrophys. 2004, 421, 13

\maketitle

\section{Introduction}

The galactic black hole transient A0620-00 is not particularly well
known for its activity in radio wavelengths.  It was last reported to have 
radio outbursts in 1975 at 962 and 151 MHz (Davis et al. 1975; Owen et al. 1976).
A few years after this observation, Duldig et al. (1979) reported a low level activity at
$2$ cm ($14.7$GHz). More recent re-analysis of the $1975$ data revealed that it underwent multiple 
jet ejection events (Kuulkers et al. 1999). There are no other reports of radio observations 
of this object.

A0620-00 is in a binary system and its mass is estimated to be around $10M_\odot$ (Gelino, Harrison \& Orosz 2001).
It was discovered in 1975 through the Ariel V sky survey (Elvis et al. 1975).
This object is located at a distance of $D=1.05$kpc (Shahbaz, Naylor \& Charles 1994). 
The outbursts and quiescence are thought to be due to some form of thermal-viscous-instability 
in the accretion disk. In the quiescent state, the accretion rate becomes very low (e.g. Lasota 2001).
Assuming there is a Keplerian disk, from optical and X-ray observations the accretion rate 
was estimated to vary from a few times the Eddington rate in outbursts 
to less than $ 10^{-11} M_\odot$ yr$^{-1}$ in quiescence (de Kool 1988; McClintock \& Remillard 1986). 
Assuming a low-efficiency flow model, McClintock \& Remillard (2000), obtained the accretion rate to be $\sim 10^{-10} 
M_\odot$ yr$^{-1}$ using X-ray observations.  A0620-00 has been in a quiescent 
state for quite some time. Our understanding of the accretion 
processes at low accretion rates suggests that magnetic field may be entangled with hot ions
at virial temperatures and could be sheared and amplified to the local equipartition
value (Rees 1984). If so, dissipation of this field, albeit small, should produce 
micro-flares from time to time, and they could be detectable especially if the object is located nearby.

In the present Paper, we report the observation of a micro-flare in radio wavelength (frequency 
1.28MHz) coming from this object. 
In the next Section, we present the details of the observations and the results.
In \S 3, we analyze our observation and compute the accretion rate in the quiescent state. 

\section{Observations and results}

On Sept. 29th, 2002, during UT 00:45-02:03 we observed A0620-00 with the Giant Meter Radio 
Telescope (GMRT) located in Pune, India. GMRT has $30$ parabolic reflector antennae placed with altazimuth
mounts each of which is of $45$ meter diameter. During our observation, 28 out of 30 antennae
were working and the observational conditions were stable. It has a tracking and pointing accuracy of $1'$
for wind speeds less than $20$km/s.
GMRT is capable of observing at six frequencies from 151MHz to 1420MHz.
On the higher side, 608-614MHz and 1400-1420MHz are protected frequency bands 
by the International Telecommunication Union (ITU).
The observed frequency is $\nu_{obs} \sim 1280$MHz far away from the ITU bands.
The band width is $16$MHz. There were 128 channels with  a channel separation of $125$kHz. 
The light-curve without the background subtraction is shown in Fig. 1. 
The data is saved every 16 seconds. The background is 
due to two side lobes and is found to be constant in time. 
The UV coverage was very good and the background was found to be constant
within the field of view with RMS noise $8.6 \times 10^{-4}$ Jy as tested by the task IMAGR in AIPS.
The UVRANGE task required no constraint. The background subtraction reveals that a micro-flare 
of average flux density of $F_\nu=3.84$mJy occurred and it lasted for
about $t_{\mu f} = 192 \pm 32$ seconds. 
We used the 3C147 as the flux calibrator and 0521+166 as the phase calibrator. 
No other source was found within the field of view. The primary beam width is $0.5$ degree
and the synthesized beam width is $3$ arc second.
The confirmation of this micro-flare is shown by the
fact that each of the antennae independently showed this event
and the synthesized image of the field of view showed no significant signal.

\begin{figure}[tb]
\vskip 5cm
\includegraphics{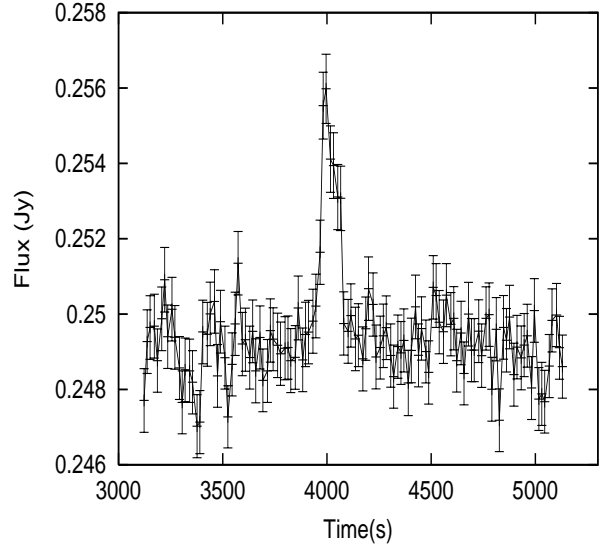}
\vskip 3cm
\begin{verse}
\caption{ Lightcurve of A0620-00 without background subtraction on Sept. 29th, 2002 as obtained
by GMRT radio observation at 1.28GHz. Subtracting the background reveals a
micro-flare of mean flux $3.84$mJy of duration $192\pm32$s. }
\end{verse}
\end{figure}

\section {Discussion}

Close to a  black hole, fast variabilities in time scales of the order of the light 
crossing time $t_l = r_g/c \sim 0.1\frac{M}{10M_\odot}$ ms, $r_g=2GM/c$ is the 
Schwarzschild radius, is possible. Shot noise in this time scale is 
observed during X-ray observations. The duration $t_{\mu f}$ of the micro-flare that we observe
is much larger ($t_{\mu f}>>t_l$), and hence we rule out the possibility that it is a shot noise type event. 

Assuming that the flare is due to magnetic dissipation, with an energy density of $B^2/8\pi$, the expression 
for the total energy release (fluence) is:
$$
E_{mag} = \frac{B^2}{8\pi} V = 4 \pi D^2 \nu_{obs} F_\nu t_{\mu f} ,
\eqno{(1)}
$$
where $V \sim r_g^3$ is the lower limit of the volume in the 
accretion flow that released the energy, $D$ is the distance of the 
source from us, $\nu_{obs}$ is the frequency at which the observation is made and
$F_\nu$ is the specific intensity of radiation. Here, $B$ is the 
average magnetic field in the inflow where the flare forms. Re-writing Eq. (1) using the equipartition law,
$$
 \frac{B^2}{8\pi} \sim \frac{GM\rho}{r} =\frac{GM{\dot M}}{4 \pi v r^3} 
\eqno{(2)}
$$ 
where $\rho$ is the density of the flow in the accretion flow, ${\dot M}$ is the 
accretion rate and $v$ is the velocity of inflow. Since there is no signature
of a Keplerian disk in the quiescent state, one may assume the inflow to be generally 
like a Bondi flow (Chakrabarti, 1990), especially close to the 
black hole. Estimations of McClintock \& Remillard (2000) 
was carried out with a low-efficiency radial flow model. Thus we use the definition of the 
accretion rate to be ${\dot M}=4\pi \rho r^2 v$.
More specifically, we assume, the free-fall velocity,  $v \sim (2GM/r)^{1/2}$. Introduction of 
pressure and rotation effects  do not change the result since the gas is tenuous,
and since the Keplerian flow is absent, the angular momentum is very low. These 
simple but realistic assumptions allow us to obtain the upper limit of the 
accretion rate of the flow to be
$$
{\dot M} \sim (3.5 \pm 0.58) \times 10^{14} x^{5/2} {\rm \ gm/s} 
= (5.5 \pm 0.91) \times 10^{-12} x^{5/2} M_\odot {\rm yr}^{-1}.
\eqno{(3)}
$$
Here $x=\frac{r}{r_g}$, is the dimensionless distance of the flaring 
region from the center. From transonic flow   
models (Chakrabarti 1990), the flow is expected to be supersonic only 
around $x\sim 2-3$ before disappearing into the black hole. 
This gives the accretion rate of A0620-00 in the quiescent state to be
$$
{\dot M}  =  (8.5 \pm 1.4) \times 10^{-11} (\frac{x}{3})^{5/2}) M_\odot {\rm yr}^{-1}.
\eqno{(4)}
$$
This is consistent with that reported by McClintock \& Remillard (2000) on the basis of X-ray observations.
It is to be noted that Duldig et al. (1979) found a flux of $44\pm14$mJy well after the 
outburst and concluded that intermittent emissions are possible and that mass transfer 
continues even in quiescence states. Our result also verifies such an assertion.

The procedure we have suggested here is general and should be applicable to 
determine the mass of the black holes if the distance is reasonably well known. Only condition
is that a hot, sub-Keplerian  component should be present.

\acknowledgements We thank the staff of the GMRT who have helped us to make this observation 
possible. GMRT is run by the National Centre for Radio Astrophysics of the 
Tata Institute of Fundamental Research. SP thanks a CSIR Fellowship which supported his 
work at the Centre for Space Physics.

{}


\begin{thebibliography}{}
\def\ref#1\par{\parshape=2 0in 14.5cm 1cm 13.5cm {#1} \par}
\parskip=0pt
\parindent=0pt

\bibitem[]{}
Chakrabarti S.K. 1990, Theory of Transonic Astrophysical Flows, (World Scientific:Singapore)

\bibitem[]{}
Davis, R.J., Edwards, M.R., Morison, I., Spencer, R.E., 1975, Nat. 257, 659

\bibitem[]{}
de Kool, M., 1988, ApJ 334 336

\bibitem[]{}
Duldig, M.L. et al., 1979, MNRAS, 187, 567

\bibitem[]{}
Elvis, M., Page, C.G., Pounds, K.A., Ricketts, M.J. and Turner, M.J.L. 1975, Nat. 257, 656

\bibitem[]{}
Gelino, D.M., Harrison, T.E. and Orosz, J.A., 2001, AJ, 122, 2668

\bibitem[]{}
Kuulkers, E., Fender, R.P., Spencer, R.E., Davis, R.J. and Morison, I., 1999, MNRAS 306, 919

\bibitem[]{}
Lasota, J.-P., 2001, New AR, 45, 449

\bibitem[]{}
McClintock, J.E., Petro, L.D., Remillard, R.A., Ricker, G.R., 1983, ApJ 266, L27

\bibitem[]{}
McClintock, J.E., Remillard, R.A., 1986, ApJ 308, 110

\bibitem[]{}
Owen, F.N., Balonek, T.J., Dickey, J., Terzian, Y., Gottesman, S.T. 1976, ApJ 203, L15

\bibitem[]{}
McClintock, J.E., Remillard, R.A., 2000, ApJ 531, 956

\bibitem[]{}
Rees, M., 1984, ARAA, 22, 471

\bibitem[]{}
Shahbaz, T., Naylor T., Charles, P.A., 1994, MNRAS 268, 756

\end{thebibliography}
\end{document}